# Ultrabroad-Band, Greatly Enhanced Light Absorption by Monolayer Graphene

**Wangshi Zhao, Kaifeng Shi, and Zhaolin Lu***

*Microsystems Engineering, Kate Gleason College of Engineering,
Rochester Institute of Technology, Rochester, New York 14623, USA*

*Corresponding author: zhaolin.lu@rit.edu

**Abstract:** We demonstrate greatly enhanced light absorption by monolayer graphene over a broad spectral range, from visible to near infrared, based on the attenuated total reflection. In the experiment, graphene is sandwiched between two dielectric media referred as superstrate and substrate. Based on numerical calculation and experimental results, the closer the refractive indices of the superstrate and the substrate, the higher the absorption of graphene will be. The light absorption of monolayer graphene up to 42.7% is experimentally achieved.

Since it was produced by mechanical exfoliation[1,2], many extraordinary properties of graphene have been discovered. For example, graphene can support remarkably high density of electric currents[3], and has high thermal conductivity[4] and elasticity[5]. At room temperature, the electron mobility of graphene is extremely high3,[6]. This enables graphene a desirable material in the advancement of nanoelectronics[7], for example, metal-oxide-semiconductor field effect transistor (MOSFET) channels[8], graphene nanoribbons[9-11], bilayer graphene transistor[12] and perforated graphene transistors[13-15]. The optical properties of graphene have intrigued considerable interest as well [16]. Recent research revealed gate-variable optical conductivity[17] and high-speed operation[18] of graphene. These extraordinary properties combining with its high electron mobility make graphene a promising candidate satisfying the need of broadband optical modulators[19-22] and photodetectors[17,23].

Considered its single atom thickness, the interaction between light and graphene is quite strong: light absorption can go up to $\pi\alpha$ = 2.293% ($\alpha$ is the fine-structure constant)[21-26] when light is normal incident through graphene. However, the absolute value (~2.3%) of absorption is still weak for most practical applications. For active optoelectronic devices[18,27,28], a strong light-matter interaction is usually desired. Therefore, many approaches have been explored to increase the interaction of light with graphene or to enhance the optical absorption. One possible way is to utilize plasmonic nanosctructures[28] or nanoparticles[29] in graphene-based photodetectors, where the responsivity can be significantly enhanced due to the localized surface plasmons. It was also shown that the enhanced absorption of graphene can be achieved by patterning doped-graphene into a periodic nanodisk[30] or alternating with insulator layers to form superlattice structure[31]. More recent research work demonstrated that over 60% absorption can be reached by integrating graphene with a microcavity structure[23]. However, all these methods mentioned above either need complicated, time-consuming fabrication processes[23,32] or the devices exhibit very narrow bandwidth[23,29,32] due to the involvement of microcavities or resonators. An easily fabricated graphene-based device with a broad bandwidth is then desired for fundamental research and practical applications.

Herein, we show that the optical absorption of monolayer graphene can be greatly enhanced over a broad spectral range, from visible to near infrared, when the incoming light with a suitable oblique incident angle based on an attenuated total reflection (ATR) configuration. This configuration has been used to measure the graphene absorption spectra[33], analyze terahertz surface plasmons on graphene[34], estimate number of carbon layers in an unknown graphene sample[35], and more recently enhance coherent light absorption by graphene[36]. In Ref. 36, up to 10% light absorption by a monolayer graphene was demonstrated through an F2 prism coupling into a graphene-sandwiched silica waveguiding structure. In our work, we found that (1) the waveguiding structure is not necessary; (2) absorptance nearly 100% can be theoretically achieved if monolayer graphene is sandwiched in a specially designed structure; and (3) the enhanced absorption can be achieved within an ultrabroad band.

In particular, we consider the graphene-sandwiched three-layer structure as shown in the inset of Fig. 1(a). The top layer (superstrate), graphene, and bottom layer (substrate) have refractive indices, $n_1$, $n_2$, and $n_3$, respectively. Assume that a plane wave is incident into the three-layer structure and the corresponding propagation angles to the normal are $\theta_1$, $\theta_2$, and $\theta_3$, respectively. Snell's law is held between layers, $n_q \sin \theta_q = n_1 \sin \theta_1$ ($q$=1, 2, 3). Based on the transfer matrix method[37], the amplitude reflectance can be calculated by

$$r = \frac{(\tilde{n}_3 - \tilde{n}_2)(\tilde{n}_2 + \tilde{n}_1)e^{+j\tilde{\varphi}_2} + (\tilde{n}_3 + \tilde{n}_2)(\tilde{n}_2 - \tilde{n}_1)e^{-j\tilde{\varphi}_2}}{(\tilde{n}_3 - \tilde{n}_2)(\tilde{n}_2 - \tilde{n}_1)e^{+j\tilde{\varphi}_2} + (\tilde{n}_3 + \tilde{n}_2)(\tilde{n}_2 + \tilde{n}_1)e^{-j\tilde{\varphi}_2}} \qquad (1)$$

where $\tilde{\varphi}_2 = n_2 k_0 d \cos\theta_2$ ($k_0$ is the wavenumber of the light wave in free space; $d \approx 0.335$nm is the thickness of graphene). Also, $\tilde{n}_q = n_q \cos\theta_q$ ($q$=1, 2, 3) for s-polarized light, and $\tilde{n}_q = n_q / \cos\theta_q$ ($q$=1, 2, 3) for p-polarized light.

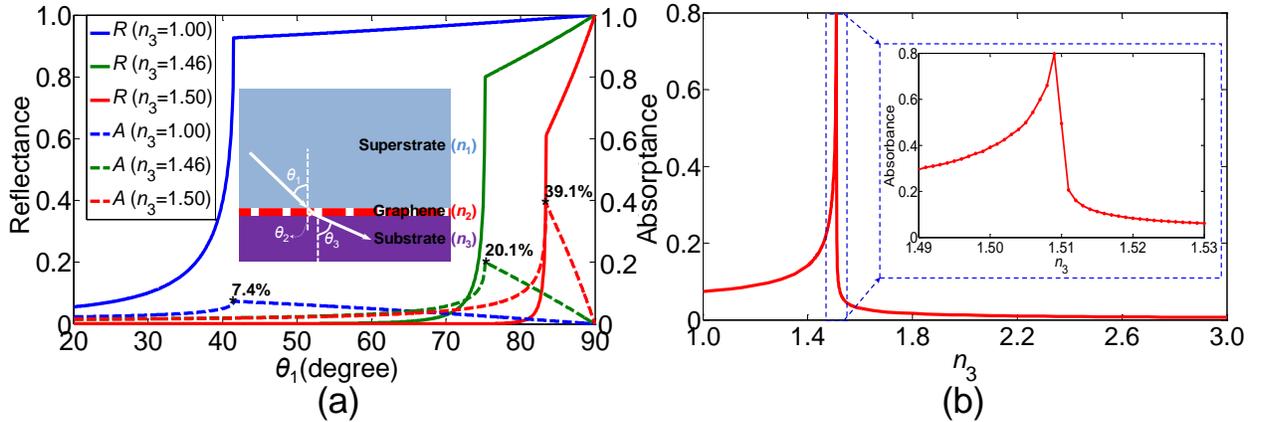

Figure 1. (a) Numerical calculation of the reflectance and absorptance as functions of incident angle, $\theta_1$, and substrate refractive index $n_3$. (b) Calculation of maximum achievable absorptance as a function of substrate refractive index $n_3$. Results in (a) and (b) are obtained at $\lambda$=650nm for s-polarized light, and $n_1$=1.51.

For pristine graphene, its surface conductivity $\sigma_g$ can be universally expressed as $\sigma_g = \frac{\pi e^2}{2h} = 6.084 \times 10^{-5}$S, where $h$ is the plank constant and $e$ represents the elementary charge. Thus, the refractive index of graphene is a complex number and can be calculated by $n_2 = \sqrt{1 + j\frac{\sigma_g}{\varepsilon_0 \omega d}}$, where $\varepsilon_0$ is the permittivity of the free space, and $\omega$ is the angular frequency of the incident light wave.

We assume the refractive index of the superstrate is $n_1=1.51$ at $\lambda=650$nm, and first consider the *s*-polarized light case. The solid lines in Fig. 1(a) plot the power reflectance $R=|r|^2$ as a function of incident angle $\theta_1$ and substrate refractive index. Based on the transfer matrix method[37], we can similarly calculate the power transmittance *T*. The absorptance $A=1-R-T$. The dashed lines in Fig. 1(a) plot the $A$-$\theta_1$ relations for different substrates. Almost identical curves can be obtained at $\lambda=1520$nm or even longer wavelengths if $n_1$ and $n_3$ remain the same as their corresponding values.

From the graph, we can see that the maximum absorptance occurs at the critical angle $\theta_c = \sin^{-1}\left(\frac{n_3}{n_1}\right)$ for each case. This can be explained as follows. Due to the ultrathin thickness of graphene, the three-layer structure can also be approximately treated as a two-layer ($n_1|n_3$) structure, with the boundary replaced by graphene as a perturbation, if the graphene absorption is not significant. According to the continuity relation of electric field in different layers, the electric field in graphene, $E_2$, is proportional to the amplitude transmittance between the superstrate and substrate

$$E_2 = t_s E_i = \frac{2n_1 \cos\theta_1 E_i}{n_1 \cos\theta_1 + n_3 \cos\theta_3}, \quad (2)$$

where $E_i$ is the electric field of the incident light wave. The power dissipation in graphene in a unit area can be calculated by $p_d = \frac{1}{2}\sigma_g |E_2|^2$, and maximum $|E_2|$ gives rise to maximum $p_d$. In Eq. (2), $|E_2|$ reaches its maximum, $2|E_i|$, when $\cos\theta_3 = 0$ or equivalently $\theta_1 = \theta_c = \sin^{-1}\left(\frac{n_3}{n_1}\right)$.

Furthermore, the closer $n_3$ and $n_1$ are, the larger the critical angle, and the larger the absorptance. The reason is because a larger critical angle decreases the incident power density on graphene, $S_i = \frac{1}{2\eta_1}|E_i|^2 \cos\theta_1$, which will increase the ratio between dissipated power and incident power, i.e.

$$A \approx \frac{p_d}{S_i} = \frac{\sigma_g \eta_1}{\cos\theta_1}\left|\frac{2n_1 \cos\theta_1}{n_1 \cos\theta_1 + n_3 \cos\theta_3}\right|^2, \quad (3)$$

where $\eta_1 = \eta_0/n_1$ is the impedance of the superstrate and $\eta_0=120\pi(\Omega)$ (the impedance of the free space). For normal light incidence into graphene suspended in air, $A \approx \sigma_g \eta_0 = 2.29\%$ in Eq. (3). When the absorption becomes significantly large, the amplitude transmittance cannot be estimated based on the above equation. Instead, the three-layer model needs to be applied.

Based on the three-layer model, the absorptance grows with the increase of $n_3$ if $n_3<n_1$, as shown in Fig. 1(b). The closer $n_3$ to $n_1^{(-)}$, the larger the absorptance is. For example, the absorptance can reach 79.6% when $n_3=1.509$. However, when $n_3>n_1$, the maximum absorptance will sharply drop. In our work, we only consider the cases where $n_1 > n_3$. The power transmittance $T=0$ when $\theta_1 \geq \theta_c$; thus, the incident power will be either reflected back or absorbed by graphene, i.e. $A=1-R$. The scattering by graphene is negligible as can be seen in the experimental result for p-polarized light. In other words, when $\theta_1 \geq \theta_c$, the absorptance can be easily measured by testing the reflectance.

Furthermore, there is no cavity or resonant component involved in the structure. As a result, the absorptance expression given in Eq. (2) is not an explicit function of frequency, which implies that the same level of enhanced light absorption can be achieved in a broad band. This is verified by the more accurate calculation based on the three-layer model: the absorptance would slightly increase from 39.1% at 650nm to 39.2% at 1620nm if there were no material dispersion ($n_1$ and $n_2$ are actually slowly varying functions of frequency due to material dispersion). Consequently, greatly enhanced ultrabroad band light absorption can be achieved simply based on the graphene-sandwiched three-layer structure.

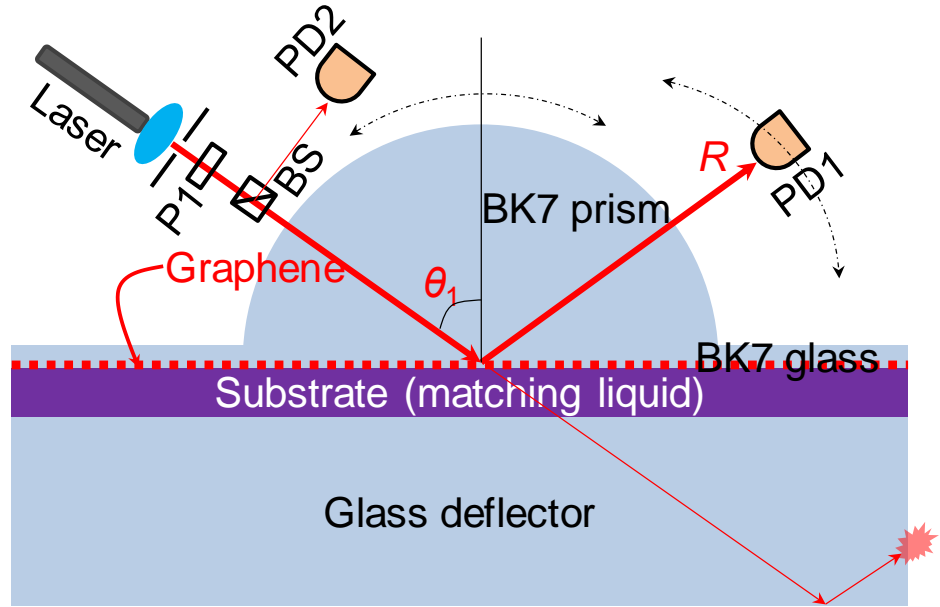

Figure 2. Illustration of the experimental setup for ATR measurement. The red dashed line represents the monolayer graphene film. P1 and BS represent the polarizer and beam splitter, respectively.

To experimentally demonstrate the greatly enhanced light absorption by graphene, we have built an ATR setup in the Kretschmann configuration[38], as illustrated in Fig. 2. A collimated laser beam propagates through a broadband polarizer (P1) to choose either s- or p- polarized light, and is then split into two by a beam splitter (BS). One beam is used for recording source power fluctuation and fed into a germanium photodiode (PD2); another is incident at the angle $\theta_1$ into a BK7 glass hemicylindrical (Ø100mm, obtained from Rocoptonics) lens, which functions as a coupling prism in this setup. The reflected light is then collected by another germanium photodiode (PD1). The power ratio between PD1 and PD2 can well measure the reflectance, R, even if there is power fluctuation in the laser source. The rotation of the prism (together with the graphene sample) and PD1 is in a $\theta_1$-$2\theta_1$ configuration, which is precisely controlled by two motorized rotation stages. In our experiment, we made one measurement for every 0.25° increment of $\theta_1$. As a result, the reflectance, R, as a function of $\theta_1$ can be plotted.

Our sample is commercially available monolayer graphene[39] synthesized by the chemical vapor deposition (CVD) process then transferred to our bare BK7 glass slide. Its Raman spectroscopy result indicates that the sample is monolayer graphene with an obviously higher 2D peak than the G peak[40]. The graphene sample is mounted at the back of the hemicylindrical prism. To avoid a thin air gap between the prism and the graphene sample, a BK7 index matching liquid is applied between them. In this case, the superstrate can be simplified and treated as a BK7 medium, consisting of the BK7 prism and the BK7 glass slide. The medium on the other side of the graphene is referred as the substrate, which is another matching liquid from Cargille$^{TM}$ in our case to better control the refractive index. A thick (>10mm) glass plate is used to hold the matching liquid and meanwhile to deflect light away from PD1.

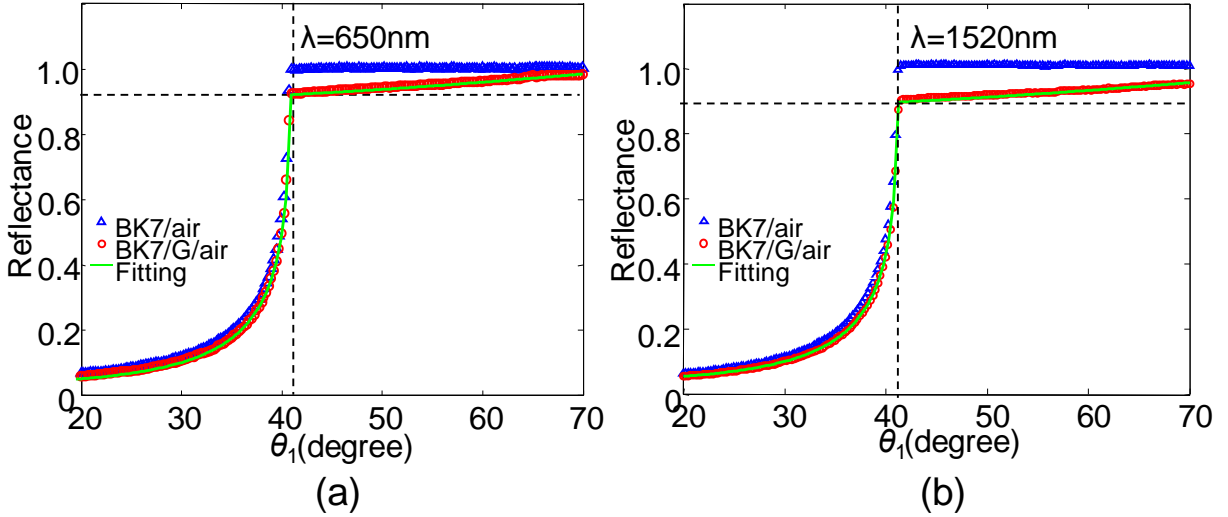

Figure 3. Reflectance of a reference BK7 glass slide (blue triangles), monolayer graphene (red circles), and numerical fit (green) with an s-polarized light (a) with wavelength of $\lambda$=650nm and (b) $\lambda$=1520nm. In the legend, "G" represents "graphene". "BK7/G/air" means the result with the BK7(prism)-graphene-air configuration.

In our work, we focused on *s*-polarization and carried out the experiment in a step-by-step fashion. First, we tested the graphene on BK7 glass sample without any substrate (or air as the substrate), as shown in Fig. 2 but without the deflector. The experimentally measured reflectance at $\lambda$=650nm and 1520nm as a function of $\theta_1$ is shown in Fig. 3(a) and (b), respectively. In order to calculate the absorptance of the monolayer graphene, the reflectance (blue triangles) of a bare BK7 glass slide is measured as a control experiment. The critical angle is measured to be 41.00° (at $\lambda$=650nm) and 41.25° (at $\lambda$=1520nm), which is in agreement with the theoretical calculation value 41.3° (at $\lambda$=650nm) and 41.8° (at $\lambda$=1520nm). The discrepancy is attributed to the angle error in $\theta_1$, and corrected in our following calculation.

In each plot, both the reflectance curves are normalized by the average value of the total internal reflection part in the reference curve. Therefore, the red curve represents the reflectance of the monolayer graphene. At the critical angle, the absorptance of the monolayer graphene can be calculated as *A*=1-*R*, which is 7.6% (at $\lambda$=650nm) and 9.8% (at $\lambda$=1520nm). The measured reflectance of the monolayer graphene is numerically fitted (green curve in Fig. 3(b)) by calculating the reflectance through the three-layer structure based on the transfer matrix method[37]. In the numerical fitting, the BK-7 medium has a refractive index of *n*=1.50 at $\lambda$=1520nm. We use the dielectric constant of graphene (imaginary part) as the fitting parameter to approach the measured reflectance data. The result turns out that the dielectric constant of graphene $\varepsilon_g$= 1+$j$15.96 or surface conductivity $\sigma_g$=6.22×10$^{-5}$S at $\lambda$=1520nm. We attribute the discrepancy between the fitted value and the theoretical value calculated by Eq. (1) to the doping in graphene. A slightly smaller conductivity can also be fitted for the result measured at $\lambda$=650nm.

Based on the same sample and configuration, we also measured the absorptance of graphene at longer wavelengths up to 1620nm. At the critical angle, the absorptance of monolayer graphene is in the range of 7.6%-11.2%, which is 3~5 times stronger than the widely known absorption coefficient (~2.3%). When the incident angle is larger than the critical angle, the reflectance of graphene is observed to gradually increase and projected to be 100% at $\theta_1 = 90°$.

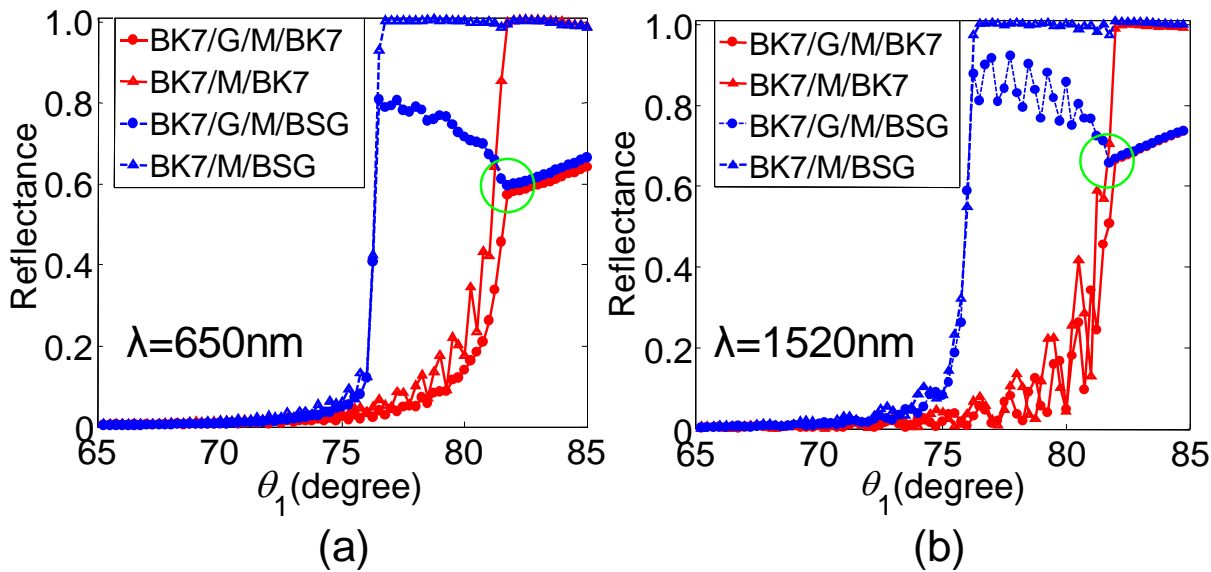

Figure 4. (a) Measured reflectance of the monolayer graphene with M1.50 as substrate under *s*-polarized light incidence at $\lambda$=650nm. (b) Measured reflectance of the monolayer graphene with M1.50 as substrate under *s*-polarized light incidence at $\lambda$=1520nm. In both graphs, red curves represent the results when a BK7 deflector is used; blue curves represent the results when a BSG deflector is used. In the legend, "G" and "M" represent "graphene" and "matching liquid", respectively. "BK7/G/M/BK7" means the result for the BK7(prism)-graphene-M1.50(substrate)-BK7(deflector) configuration.

The absorption of the monolayer graphene can be further enhanced when the refractive index of the substrate increases. In our work, a matching liquid with refractive index 1.50 (at $\lambda$=589.3nm according to the manufacturer; referred as "M1.50") is applied as the substrate of the graphene. Two different supporting glass deflectors, BK7 and borosilicate glass (BSG), are separately used to hold the substrate M1.50. In Fig. 4, triangle points are the control experiments measured by "removing" the graphene. Both the curves measured with graphene sample are normalized with the corresponding control curves.

In the control experiment with BK7 deflector, the total internal reflection occurs at the interface of the superstrate BK7 and the substrate M1.50, $\theta_C \approx 81.75°$. As shown in the red dot curve, at $\theta_C$ the absorptance of monolayer graphene is measured as 42.7%, which is ~18 times stronger than the widely known absorptance (~2.3%)!

This result is further confirmed by replacing the BK7 deflector with the BSG deflector and repeating the measurement, as shown in the blue dot curves in Fig. 4 (a), where the absorption is measure as 40.5%. In this configuration, there are two total internal reflections: first one occurs at the interface of the substrate M1.50 and BSG deflector with a critical angle $\theta_{C1} \approx 76.75°$, and the second one occurs at the interface of the superstrate BK7 and the substrate M1.50 with $\theta_{C2} \approx 81.75°$. Beyond $\theta_{C2}$, the blue dot curve is in a good agreement with the red dot curve, and both indicate the reflectance of the monolayer graphene in the sandwich configuration. The oscillation of the blue triangle curve between the two angles $\theta_{C1}$ and $\theta_{C2}$ is due to the substrate M1.50 functioning as a cavity between the graphene and BSG deflector.

Similarly, $R$-$\theta_1$ relation at different wavelengths is measured. Figure 4(b) shows the results for the measurement at $\lambda$=1520nm. The absorptance is about 35.3% at the critical angle. When the wavelength varies from 650nm to 1620nm, the absorptance gradually decreases from 40.5% to 33.1% with the BSG deflector, and from 42.7% to 35.3% with the BK7 deflector. Thus, ultrabroad band enhanced light absorption is achieved. The variation of the absorptance with the wavelength can be attributed to the dispersion of the superstrate BK7 and substrate M1.50. In particular, the absorptance becomes more sensitive when $n_1$ and $n_3$ are close enough. When the substrate is replaced by M1.49 matching liquid, the absorptance will considerably decrease, for example, dropping from 42.7% to 28% at $\lambda$=650nm.

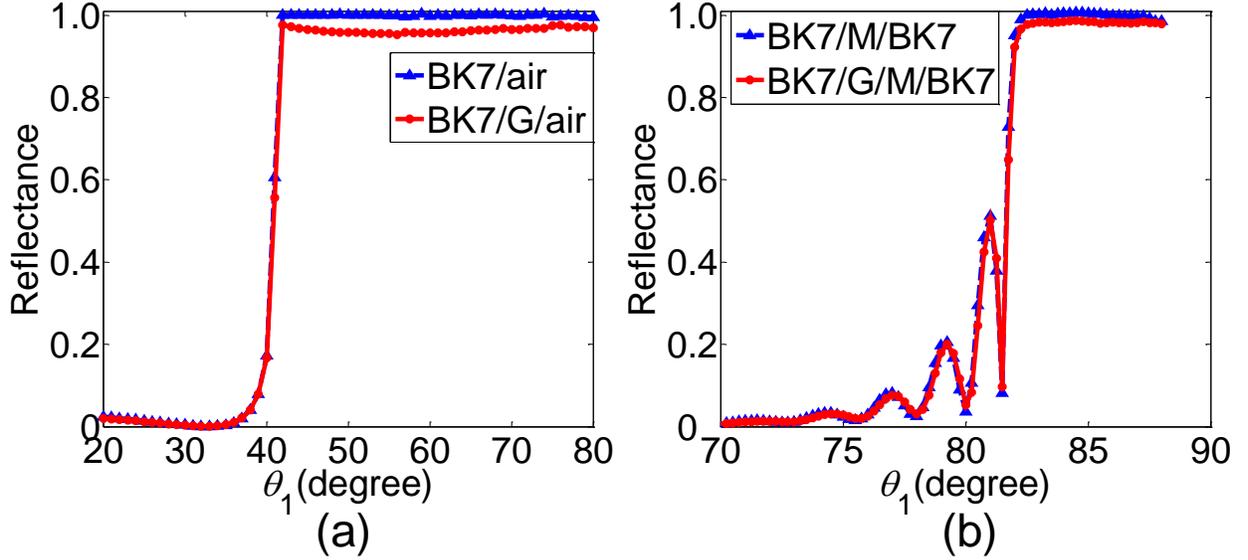

Figure 5. For 1520nm $p$-polarized light incidence, the measured reflectance by the monolayer graphene sample as a function of incident angle when (a) the substrate is air, and (b) the substrate is M1.50.

The enhanced absorption of graphene is shown to be very sensitive to the polarization of the incident light. Similar as previous experiments, we measured the reflectance of the monolayer graphene when the substrate is air and M1.50 separately with $p$-polarized incident light. As can be seen in Fig. 5, the maximum absorptance does not occur at the critical angle and is not sensitive to the substrate refractive index. The minimum reflectance beyond the corresponding critical angle is 97.65% (when the substrate is air) or 97.7% (when the substrate is M1.50), respectively. Therefore, the absorptance in both cases is calculated as ~2.3%, which is similar as the widely known absorption coefficient. The reason is because the electric field in graphene for $p$-polarized incident light cannot be estimated by Eq. (2), instead

$$|E_2| = \sqrt{|E_{2t}|^2 + |E_{2n}|^2} \approx \left|\frac{2n_1 \cos\theta_1 E_i}{n_1 \cos\theta_3 + n_3 \cos\theta_1}\right| \sqrt{\left(|\cos\theta_3|^2 + \left|\frac{n_3}{n_2}\sin\theta_3\right|^2\right)}. \quad (4)$$

where $E_{2t}$ and $E_{2n}$ represents the tangential and normal components of the electric field in graphene. Therefore, when $\theta_1 = \theta_C = \sin^{-1}\left(\frac{n_3}{n_1}\right)$, $\cos\theta_3 = 0$ and $E_{2t} = 0$, which is opposite to that case for *s*-polarized light. In the latter case, $|E_2| = |E_{2t}|$ gains its maximum when $\theta_1 = \theta_C$.

To summarize, we have experimentally demonstrated that the absorption of monolayer graphene can be significantly enhanced over a broad spectral range, from visible to infrared, when the incident light is *s*-polarized. At the critical angle, the absorption is in the range of 7.6%-11.2%, when the substrate is air, and up to 42.7%, when the substrate is replaced by a medium with a closer refractive index compared to that of the superstrate. The enhanced absorption is not strongly dependent on the wavelength but very sensitive to the polarization of incoming light. The significantly enhanced absorption of monolayer graphene may have potential applications in broadband photodetectors and solar cells.

Although all investigation in this report is focused on monolayer graphene, the absorption by bilayer, 3-layer, 4-layer, 5-layer graphene can be enhanced to 73%, 79%, 88%, and 93%, respectively based on the same configuration.

This material is based upon work supported by the National Science Foundation under Award No. ECCS-1308197 and ECCS-1057381. Acknowledgement is made to the Donors of the American Chemical Society Petroleum Research Fund for partial support of this research.